\documentclass[aps,prl,twocolumn,showpacs,superscriptaddress,floatfix,nofootinbib,amsmath,amssymb,eufrak,table,
]{revtex4-2}
\usepackage{hyperref}
\usepackage{amsmath}
\usepackage{bm}
\usepackage{color}
\usepackage{graphicx}

\definecolor{Tomato2}{rgb}{0.932,0.36, 0.26}
\definecolor{RoyalBlue2}{rgb}{  0.264, 0.43,  0.932 }
\definecolor{abhijit}{rgb}{1,0.4,0.8}

\hypersetup{
colorlinks=true,
pdfstartview=FitV, 
linkcolor=RoyalBlue2,
citecolor=Tomato2,
urlcolor=Tomato2
}

\begin{document}

\title{Bulk medium evolution has considerable effects on jet observables}

\author{Yasuki~Tachibana} 
\affiliation{Department of Physics and Astronomy, Wayne State University, Detroit, Michigan 48201, USA}
\affiliation{
Akita International University, Yuwa, Akita-city 010-1292, Japan
}

\author{Chun~Shen}
\affiliation{Department of Physics and Astronomy, Wayne State University, Detroit, Michigan 48201, USA}
\affiliation{RIKEN BNL Research Center, Brookhaven National Laboratory, Upton, New York 11973, USA}

\author{Abhijit~Majumder} 
\affiliation{Department of Physics and Astronomy, Wayne State University, Detroit, Michigan 48201, USA}

\date{\today}

\begin{abstract} 
    We consider the case, in QCD, of a single jet propagating within a strongly interacting fluid of a finite extent. Interactions lead to the appearance of a source of energy-momentum within the fluid. The remnant jet that escapes the container is analyzed along with portions of the medium excited by the jet. We study the effect of a static versus a semi-realistic expanding medium, with jets traveling inward versus outward. 
    We consider the medium response via recoils in partonic scatterings based on a weakly coupled description and its combination with hydrodynamical medium response based on a strongly coupled description, followed by incorporation into a jet. The effects of these limits on the reconstructed energy, momentum, and mass of the jet, as a function of the angle away from the original parton direction, are studied. It is demonstrated that different flow velocity configurations in the medium produce considerable differences in jet observables. This work highlights the importance of accurate dynamical modeling of the soft medium as a foundation on which to calculate jet modification and casts skepticism on results obtained without such modeling. 
\end{abstract}

\maketitle
Over the last several years, 
the study of jets as probes~\cite{Bjorken:1982tu,Baier:1996kr,Baier:1996sk,Zakharov:1996fv,Gyulassy:1999zd,Gyulassy:2000fs,Gyulassy:2000er,Wiedemann:2000za,Wiedemann:2000tf,Guo:2000nz,Wang:2001ifa,Majumder:2009ge,Arnold:2001ba,Arnold:2002ja,Majumder:2010qh} 
has taken center stage in the exploration of the quark-gluon plasma (QGP), 
created in nuclear collisions at the BNL Relativistic Heavy-Ion Collider (RHIC) and the CERN Large Hadron Collider (LHC).
The first experimental signature of jet quenching was discovered at RHIC 
through high-$p_{T}$ hadron suppression~\cite{Adcox:2001jp,Adams:2003kv}. 
Later, modifications of reconstructed jets were observed at the LHC~\cite{Aad:2010bu,Chatrchyan:2011sx,Abelev:2013kqa}. 
Developing experimental techniques allow us to investigate 
jets and jet modification via a variety of observables: 
Inclusive observables such as 
the nuclear modification factor~\cite{Aad:2014bxa} and azimuthal anisotropy~\cite{Aad:2013sla} 
as well as substructure observables such as the jet fragmentation function~\cite{Aad:2014wha}, jet shape~\cite{Chatrchyan:2013kwa}, jet mass~\cite{Acharya:2017goa}, etc.

Most early calculations of jet quenching used the Bjorken cylinder~\cite{Gyulassy:2001kr,Vitev:2002pf,Salgado:2003gb} as the background medium: 
The cross-sectional area was given by the overlap of the two incoming nuclei,  while the local entropy density would diminish due to longitudinal expansion, as $s(\tau) = s(\tau_0)\tau_0/\tau$. 
While such a simplified medium profile was known to yield specious results for some high-$p_{T}$ hadron observables such as the azimuthal anisotropy $v_2$~\cite{Majumder:2006we,Bass:2008rv}, 
it yielded a fair description of the $p_{T}$ and centrality dependence of leading hadron suppression~\cite{Vitev:2002pf,Salgado:2003gb}.

Within the last decade there has been a near sea change in our understanding of jets, from their clustering algorithms~\cite{Cacciari:2008gp}, 
the effect of higher orders and resummations~\cite{Dasgupta:2014yra,Kang:2016mcy}, 
to the variety of minor effects that influence their modifications in a dense medium~\cite{Thoma:1990fm,Majumder:2008zg,Beraudo:2011bh,Borghini:2005em,Majumder:2007ae}. 
However, in the majority of cases, there has been little improvement in the modeling used to describe the underlying medium that both provide the space-time profile of the energy or entropy density through which the jets propagate and which the jets can in turn excite. 

The argument that, due to asymptotic freedom, a jet will be weakly coupled with the soft medium, and as such an accurate description of the bulk dynamics of the medium is \emph{not} essential for jet observables, is simply incorrect. 
Jets include both hard modes (weakly coupled with the medium) and a multitude of soft modes, branched off from the hard modes. 
These softer modes, which start off as soft partons, interact strongly with the medium constituents, leading to the excitation of the medium.
Depending on its energy, a considerable portion of the jet in a heavy-ion collision will consist of softer hadrons that originate from the excited medium.  
In this Letter, 
we point out that the bulk dynamics of the underlying medium, 
by interacting strongly with the softer modes of the jet, 
exerts an unexpected amount of influence on the properties of the modified jet. 

To illustrate our point, 
we use a state-of-the-art event generator as furnished by the JETSCAPE collaboration~\cite{Putschke:2019yrg,JETSCAPE:2019udz,JETSCAPE:2020shq,JETSCAPE:2020mzn,JETSCAPE:2021ehl, JETSCAPE:2022cob, JETSCAPE:2022jer}
~(JETSCAPE 1.0). 
The modification of the parton shower in the medium is calculated within a multistage approach, 
using the MATTER~\cite{Majumder:2013re, Cao:2017qpx} simulator for partons with virtualities $\mu\geq2$~GeV and 
the LBT~\cite{Luo:2018pto,He:2018xjv} simulator for partons with $\mu \approx 2$~GeV, in the medium. 
Virtuality refers to the off-shellness of a parton i.e., $\mu^2 = p^\mu p_\mu - m^2$.
At high virtualities, partons in MATTER 
undergo vacuum-like virtuality ordered splitting process with later emissions taking place at successively lower virtualities. 
The scattering in the medium engenders a minor enhancement in the emission rate by small temporary increments of the virtuality. 
As virtualities of the later produced partons drop down to a scale comparable to that generated by scatterings in the medium, one enters the regime of multiple scattering-induced emissions. 
This stage, modeled using the LBT generator, uses the medium modified emission kernel to calculate 
the rate of the emission, which is induced only by the medium effect. 
Once the parton escapes the QGP, it may re-enter the MATTER phase without a medium-induced portion,
if the virtuality is higher than $\mu_{\mathrm{min}}=1$~GeV. 

More details on the partonic portion of this multistage event generator may be found in Ref.~\cite{JETSCAPE:2017eso}, and will not be repeated here. 
The main difference between the calculations in that effort and the current is the value of the jet transport coefficient $\hat{q}$, 
which controls the strength of medium-induced emissions in both MATTER and LBT. 
Partons within the jet shower scatter off partons within the medium, this scattering is the source of medium-induced radiation. 
The transport coefficient $\hat{q}$ represents the mean square momentum, transverse to the velocity of the parton, exchanged between it and the medium.
We will assume a quasi-particle picture of the medium, as furnished by the hard thermal loop effective theory~\cite{Wang:2000uj,CaronHuot:2010bp}.
Given a medium temperature $T$ at the location of the parton $i$ (with energy $E_i$), one obtains: 
\begin{eqnarray}
\hat{q}_i =C_{i}\alpha^{\mathrm{med}}_{\mathrm{S}}\mu_{\mathrm{D}}^2T
\log\left(\frac{6E_i T}{\mu_{\mathrm{D}}^2}\right).
\label{qhat}
\end{eqnarray}
Here, $C_{i}=C_{\mathrm{A}}$ or $C_{\mathrm{F}}$ is the Casimir color factor for the parton $i$, $\alpha^{\mathrm{med}}_{\mathrm{S}}$ is the in-medium coupling constant, and 
$\mu_{\mathrm{D}}=gT[N_{\mathrm{c}}/3+N_{\mathrm{f}}/6]^{1/2}$ is the Debye screening mass for a QCD plasma with $N_{\mathrm{c}}= 3$ colors and $N_\mathrm{f}= 3$ fermion flavors.

In this Letter, 
we examine two scenarios for the medium response to jet propagation. 
One is based on the weakly coupled description 
in which all processes of energy-momentum exchange between jets and medium are carried out through individual partonic scatterings. 
The scatterings are simulated as 
$2$-to-$2$ processes with a jet parton and a parton picked up from the medium via sampling of the thermal distribution. 
After the scattering, the outgoing partons, called recoil partons, propagate while interacting with the medium in the same way as jet partons. 
The energy-momentum deficits (\emph{holes}) left in the medium for the thermal partons sampled in the $2$-to-$2$ interaction 
are to be subtracted from the final jet energy and momentum. 
In this scenario, the medium response to jet propagation is constructed by the propagation of recoils and their successive interactions. 

The other scenario is the strongly coupled description in which 
the transport of jet energy and momentum is extended to the hydrodynamic regime. 
In this scenario, 
the soft partons with energy below a cut $E^\mathrm{dep}_\mathrm{cut}$ (as well as the \emph{holes} generated in the $2$-to-$2$ interaction)
are assumed to be absorbed in the fluid,   
and their energy and momentum are diffused until they \emph{thermalize}. 
Then, the diffused energy and momentum are injected into the medium and evolve hydrodynamically with the bulk medium. 

The diffusion process in the strongly coupled description for the medium response 
is modeled with the causal relativistic diffusion equation~\cite{Aziz:2004qu}: 
\begin{eqnarray}
\left[
\frac{\partial}{\partial t}+\tau_\mathrm{relax}\frac{\partial^2}{\partial t^2}-D_\mathrm{diff}\nabla^2
\right]
j_i^\nu(x)
&=&
0
\label{eq:causal_diff}
\end{eqnarray} 
with the initial conditions 
$j_i^\nu = \pm_{i} \, p_i^\nu \delta^{(3)}(\vec{x}-\vec{x}^\mathrm{dep}_i)$ and $\partial j_i^\nu/\partial t =0$ at $t=t^\mathrm{dep}_i$, 
where $\pm_{i}$ corresponds to a soft parton being absorbed or a \emph{hole} corresponding to a deficit in the scattering, 
with an index $i$ and four-momentum $p_i^\nu$ at $( t^\mathrm{dep}_i, \vec{x}^\mathrm{dep}_i )$.
Here, $D_\mathrm{diff}$ is the diffusion coefficient and $\tau_\mathrm{relax}$ is the relaxation time. 
The speed of signal propagation in Eq.~(\ref{eq:causal_diff}) is given by $v_\mathrm{sig}=(D_\mathrm{diff}/\tau_\mathrm{relax})^{1/2}$ and the values of $D_\mathrm{diff}$ and $\tau_\mathrm{relax}$ are chosen to satisfy $v_\mathrm{sig}\leq 1$. 

The calculation of the diffusion process is carried out in the limit where flow velocity and the gradients of the medium fluid is much smaller than the expansion rate and gradients of the wake by the energy-momentum deposition. 
Under this assumption, the parameters controlling the process are regarded as constants. The diffusion equation is reduced to the one without the effects of anisotropy and inhomogeneity of the background medium shown in Eq.~(\ref{eq:causal_diff}). 
This approximation can break down when the medium has finer structures, such as with a more realistic fluctuating initial condition. 
To accurately study the in-medium thermalization process through more realistic simulations, parameters dependent on local thermodynamic quantities need to be formulated in more detail, and background flow effects must be treated appropriately. 
We leave such further improvement for future work. 

The equation of motion for the fluid with deposited energy and momentum is given by  
$\partial_{\mu}T^{\mu \nu}(x)=J^{\nu}(x)$ 
\cite{Tachibana:2017syd,Chen:2017zte,Chang:2019sae}, 
where $T^{\mu \nu}$ is the energy-momentum tensor of the medium. 
The source term $J^\nu$ can be written as 
$J^{\nu}(x)=\sum_i j_i^{\nu}(x) \delta ( t-[t^\mathrm{dep}_i + t_\mathrm{th}])$, 
where $j_i^{\nu}(x)$ is the solution of Eq.~(\ref{eq:causal_diff}). We assume the entire energy-momentum source term $j_i^{\nu}(x)$ thermalizes after a thermalization time $t_\mathrm{th}$. 
The medium is modeled as an ideal fluid with an equation of state (EoS) from Ref.~\cite{Borsanyi:2013bia}.

To investigate 
how the background medium expansion affects the jet shower evolution and the medium response to it, 
we employ two different initial profiles of the medium for a systematic comparison. 
One is a static uniform initial condition, the so-called brick, with infinite size and temperature $T=0.25$~GeV. 
The other is given by an expanding oblate three-dimensional Gaussian profile of energy density, 
\begin{eqnarray}
\epsilon(t\!=\!0,x)=\epsilon_\mathrm{lat}(T_0)\,
\mathrm{exp}\!\!
\left[
-\!\left(\frac{x^2\!+\!y^2}{2\sigma_{\mathrm{T}}^2}
\!+\! \frac{z^2}{2\sigma_z^2}
\right)
\right],
\label{eq:profile}
\end{eqnarray}
where $\epsilon_\mathrm{lat}$ is the energy density at the center (at a set initial temperature $T_0=0.5$~GeV). 
To mimic the medium's rapid longitudinal expansion in heavy-ion collisions, 
we set the width of the profile in the $z$-direction to be 
smaller than that in the transverse direction: $\sigma_{\mathrm{T}}=1.5$~fm and $\sigma_z=0.75$~fm.
This expanding fluid profile is different from the one used commonly in studies for heavy-ion collisions. 
However, the initial transverse size, energy density profile, as well as longitudinal and transverse expansion are similar to that of a realistic simulation (see Supplemental Material \cite{SupplMat} for a comparison between this medium and Bjorken expansion). 
This particular choice also allows us to avoid coordinates designed for collider configurations 
in which the beam axis is taken as a special direction. 

In this study, we define the jet energy, momentum, and mass squared as
\begin{eqnarray}
P^\mu(\theta)=\int_0^\theta d\theta' \frac{dP^\mu}{d\theta'},
\hspace{5pt}
M^2(\theta)=P^\mu(\theta)P_\mu(\theta), 
\label{eq:jet_mom_mass}
\end{eqnarray}
to study how these conserved quantities assigned to the initial parton are recovered by increasing the polar angle $\theta$ from the original  parton  direction. 
Throughout the letter, radians are used to measure angles.
Here, $dP^{\mu}/d\theta$ is the angular distribution of four-momentum associated with jet propagation, 
which is obtained with background subtraction appropriate for each type of medium contribution. 
In the weakly coupled case,
we employ the commonly used subtraction method 
for \emph{hole} partons \cite{Luo:2018pto},
\begin{eqnarray}
\label{eq:neg_sub}
\!\!\!\!
\frac{dP_\mathrm{weak}^\mu}{d\theta}
&=&
\frac{1}{\Delta \theta} 
\!
\left[ 
\sum\limits
_{\substack{i\in\mathrm{shower}\\
p^0_i\geq E^\mathrm{dep}_\mathrm{cut}}}
^{\theta < \theta_i < \theta + \Delta \theta}
  \!\!\!\!\!\!\! p^{\mu}_i 
+
\!\!
\sum\limits
_{\substack{i\in\mathrm{shower}\\
0\leq p^0_i< E^\mathrm{dep}_\mathrm{cut}}}
^{\theta < \theta_i < \theta + \Delta \theta}  \!\!\!\!\!\!\! p^{\mu}_i 
-
\!\!
\sum\limits_{j\in \mathrm{holes}}^{\theta < \theta_j < \theta + \Delta \theta}  \!\!\!\!\!\!\! p^{\mu}_j   
\right]\!\!,  
\end{eqnarray}
where $p^\mu_i$ and $\theta_i$ are parton $i$'s 
four-momentum and polar angle from the original parton's direction. 
In Eq.~(\ref{eq:neg_sub}), 
the sum of the first two terms on the right hand side 
is taken over all final state partons in the shower.
The sum in the last term is taken over all \emph{hole} partons. 

On the other hand, in the strongly coupled description,
some portion of jet energy and momentum propagates via hydrodynamic flow in the medium. 
The fluid part contribution 
needs to be added to the final jet, i.e.,
\begin{eqnarray}
\!\!\!\!\!\!\!
\frac{dP_{\mathrm{strong}}^\mu}{d\theta}
\!=\!\!
\frac{1}{\Delta \theta} 
\!\!
\sum\limits
_{\substack{i\in\mathrm{shower}\\
p^0_i\geq E^\mathrm{dep}_\mathrm{cut}}}
^{\theta < \theta_i < \theta + \Delta \theta}
\!\!\!\!\! p^{\mu}_i \!
+\!\!
\left.
\frac{d P^\mu_\mathrm{fluid}}{d \theta}
\right|_\mathrm{w\!/\,jet}
\!\!-\!\!
\left.
\frac{d P^\mu_\mathrm{fluid}}{d \theta}
\right|_\mathrm{w\!/\!o\,jet}\!. 
\label{eq:mom_strong}
\end{eqnarray}
The result from a hydro simulation without jet propagation is subtracted as the background. 
Note here that only partons with energy above the cut $E^\mathrm{dep}_\mathrm{cut}$ exist in the shower in this scenario. 
To obtain the fluid contribution 
$d P^\mu_\mathrm{fluid}/d \theta$, 
we need to include the effect from medium particlization, which is done via the well-known Cooper-Frye prescription~\cite{Cooper:1974mv}: 
\begin{eqnarray}
\label{eq:bs_after_fo}
\!\!\!
\frac{dP^{\mu}_\mathrm{fluid}}{d\theta}
=
\int \!\!\! dpd\phi |\vec{p}|^2 
\sin\theta\!
\left[
\sum_i \!\int_{\Sigma_\mathrm{FO}}
\!\!\!\!
\frac{p^{\alpha}d^3\!\sigma_{\alpha} }{p^0}p^\mu f_{i}(x,p)\!\right]\!,
\end{eqnarray}
where $f_i$ is the local equilibrium distribution of particles in the medium 
and the integration in square brackets is over the freeze-out hypersurface $\Sigma_\mathrm{FO}$, 
for which we use the isochronous surface at $t_\mathrm{FO}=10$~fm/$c$ for simplicity.

To see the hydrodynamic transport directly i.e., without particlization, 
we also calculate the medium contribution from the energy-momentum tensor of the fluid: 
\begin{eqnarray}
\!\!\!\!\!\!\frac{d P^\mu_\mathrm{fluid}}{d \theta}
&=&
\int d^3x 
T^{0\mu}
(t_\mathrm{FO},\vec{x})
\delta(\theta-\theta_{T}),
\label{eq:bs_before_fo}
\end{eqnarray}
where $\theta_{T}$ is the polar angle of fluid momentum $T^{0i}$ with respect to the jet direction.
For the subtracted background in the brick case, 
we take the limit of $T^{0i}\to0$ for all spatial components.

In this work, 
we perform simulations of jet evolution in a vacuum, a static brick, and an expanding medium, 
starting with a single gluon of energy $E_\mathrm{init}\!=\!140$~GeV (see Supplemental Material \cite{SupplMat} for a result for smaller initial jet energy). 
Its momentum amplitude is determined by sampling the Sudakov form factor in MATTER, with a direction along the $x$-axis.
The jet partons evolve up to $t\!=\!8$~fm. For vacuum simulations, we apply MATTER without medium effects. 
In-medium energy loss is modelled by MATTER and LBT with $\alpha^{\mathrm{med}}_\mathrm{S}=0.25$, above $T = 160$ MeV. 

In the weakly coupled case,
the medium evolves according to 
the hydrodynamic equation without a source term. 
In the strongly coupled description, 
we introduce source terms generated by the relativistic diffusion equation~(\ref{eq:causal_diff}) 
for the \emph{hole} partons and soft partons with the energy in the aboratory frame below the cut 
$E^\mathrm{dep}_\mathrm{cut}\!=\!2$\,GeV.
We set $\tau_\mathrm{relax}\!=\!1.0$\,fm, $D_\mathrm{diff}\!=\!0.6$\,fm, and $t_\mathrm{th}\!=\!1.5$\,fm.

\begin{figure*}[htbp]
\begin{center}
\includegraphics[width=0.9\textwidth]{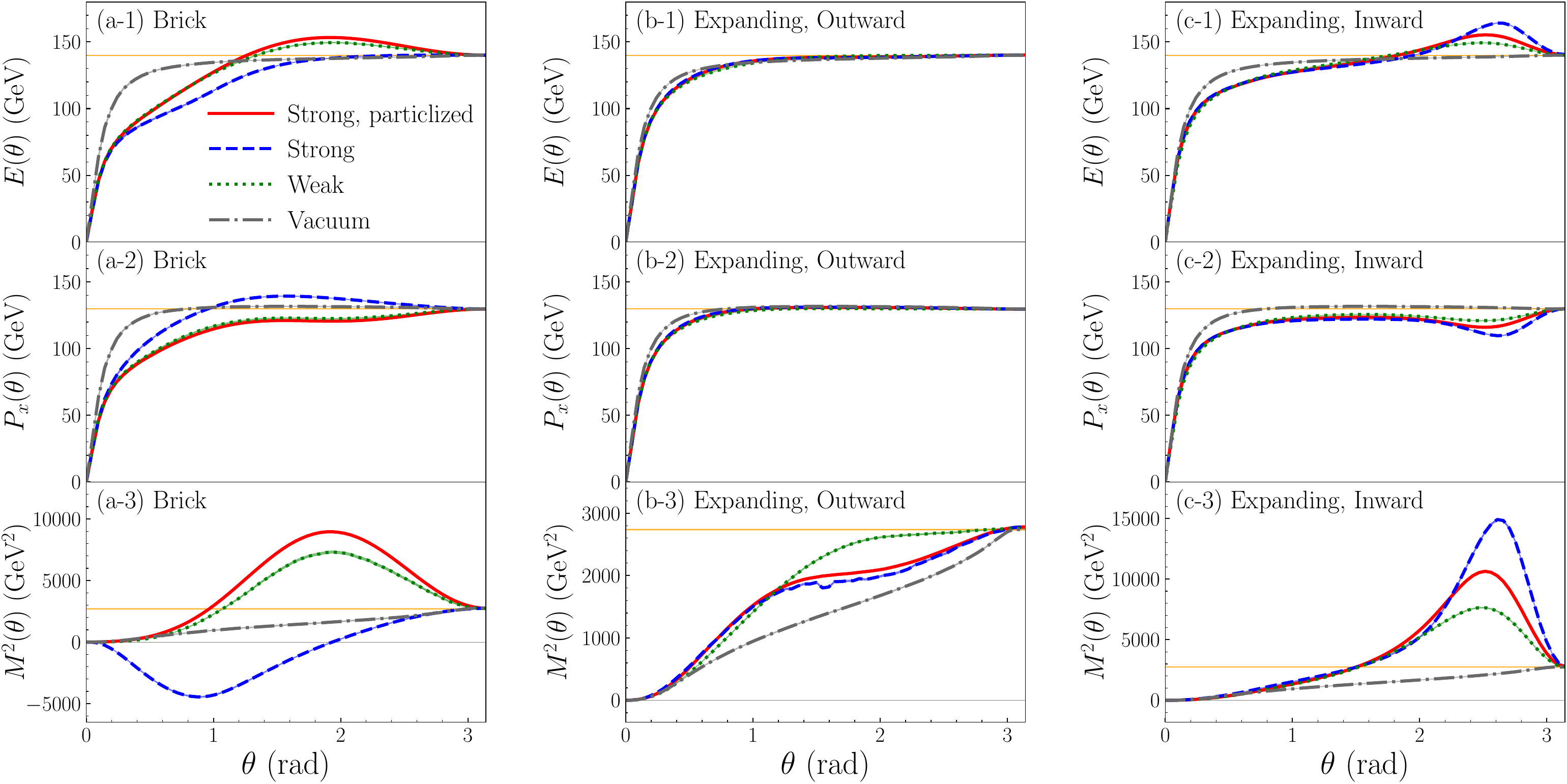}
\vspace{-20pt}
\end{center}
\caption{(Color online) 
Jet energy, momentum in the direction of the initial parton, and mass squared as a function of the opening angle $\theta$ (in radians) for the jets propagating 
in the brick medium (a-1)--(a-3), 
outward along the expanding medium (b-1)--(b-3), 
and 
inward against the expanding medium (c-1)--(c-3). 
The solid lines and dashed lines are the results including the hydrodynamic response effect in a strongly coupled description with and without particlization, respectively. 
The dotted lines show the results with the medium response within a weakly coupled description by recoils. 
The dot-dashed lines represent vacuum simulations. 
The horizontal orange lines indicate the average values assigned to the initial parton. }
\label{fig:1}
\end{figure*}

Figure~\ref{fig:1} shows the angular distribution of the jet observables defined in Eq.~(\ref{eq:jet_mom_mass}). 
For the case with the expanding medium, 
we consider two different production points of the jet initiating parton; 
the parton is produced at the center of the medium $(x\!=\!y\!=\!z\!=\!0)$ and escapes \emph{outward} along with the radial flow, 
or 
the parton is produced at the edge $(x\!=\!-10~\mbox{fm},y\!=\!z\!=\!0)$ and propagates \emph{inward}, toward the center against the radial flow. 

It should be noted that all jet energies, momenta, and masses squared, 
for all settings, converge to their original values, 
which are assigned to the jet initiating parton, 
at $\theta = \pi$, showing their conservation during the evolution. This also serves as a crucial test of our calculation. 
The jet substructures are modified within the constraints of these conservation conditions. 

\emph{In the case of the brick}: 
The results from the strongly coupled scenario without particlization (dashed-blue lines) indicate 
that the energy-momentum propagation is spread away from the initial jet direction, appearing in directions both behind and away from the jet, with the energy recovered monotonically.  
Meanwhile, the momentum increases sharply until $\theta=\pi/2$ and then very slowly decreases to its full value at $\pi$. 
Remarkably, this structure with momentum larger than energy 
gives a negative value of the mass squared over a wide range of $\theta$.

Particlization causes suppression of energy-momentum emission in the direction opposite to the jet. It drastically modifies the jet structure by flipping the order of how the net energy and momentum approach to their full values around $\theta=\pi$. This is a consequence of the subtraction of the locally boosted equilibrium distribution used in Eq.~(\ref{eq:bs_after_fo}).
The particle distribution in a fluid element with flow velocity, 
which tends to be in the jet direction caused by jet-energy deposition, gives fewer particles with momentum opposite to the flow, but more particles in the flow direction, than the isotropic distribution in a static fluid. 

The results from the weakly coupled scenario show 
qualitatively the same trends as in the strong particlized case (dotted-green lines vs. solid-red lines). 
In a static medium, the subtraction prescription for \emph{hole} partons, in Eq.~(\ref{eq:neg_sub}), can somewhat mimic the backward suppression as well as the broadening. 
Some medium partons involved in the scatterings 
can initially have momentum opposite to the jet direction 
and be bounced forward. 

\emph{In the case of jet propagation outward}: From the expanding medium, 
the strong radial flow following the jet pushes the transported energy and momentum forward. This blue shift generates a focus effect and makes the broadening of the energy and momentum response distribution more moderate than the static situation.
In this case, the particlization effect is very negligible. 
This is because 
most particles in a fluid element with a very large flow velocity tend to be aligned with the flow velocity. 
Thus, the fluid elements cannot emanate particles as opposed to the strong radial flow of the medium, regardless of the direction of the jet-induced flow. 
Therefore, the backward over subtraction 
is not as remarkable as in the brick case.
In the expanding medium case, 
the angular evolution of jet mass shows a clear difference between the weakly- and strongly coupled scenarios when $\theta > 1$.
This is because the medium partons excited in the weakly coupled scenario are sampled from the local temperature and flow velocity at each parton split. This particle distribution is different from those obtained from the Cooper-Frye procedure [Eq.~\eqref{eq:bs_after_fo}] applied to fluid cells at the switching surface for particlization.

\emph{In the case of jet propagation inward}: 
In this case, the jet-induced flow following the jet dams up and deflects off the opposing radial flow in the strongly coupled scenario. %
Thereby, the momentum radiated from the jet is pushed in the backward direction.  These deflected vectors continue to add to the energy accumulated in the backward direction. 
These structures after the particlization are smeared but still show the same tendency.
In the weakly coupled scenario, the recoil partons with multiple successive collisions can somewhat mimic the jet-induced flow, as they are sampled from the oppositely flowing medium.

\begin{figure*}[htbp]
\begin{center}
\includegraphics[width=0.9\textwidth]{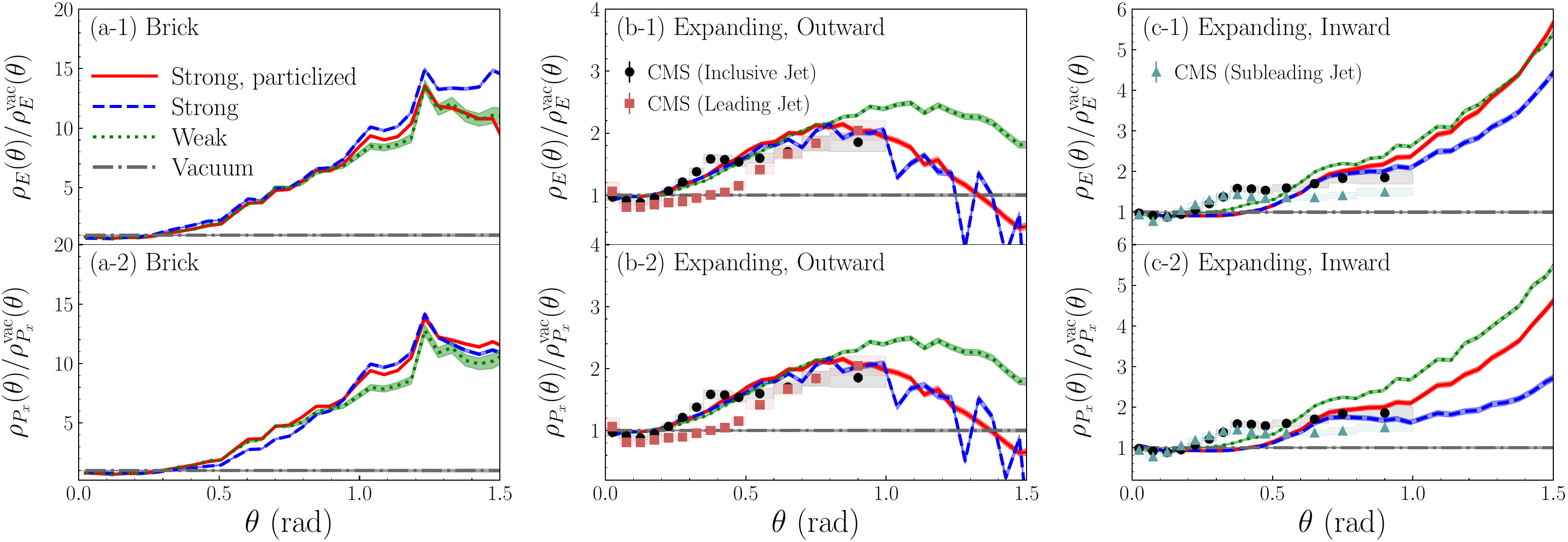}
\vspace{-20pt}
\end{center}
\caption{(Color online) 
Ratio of the jet energy and momentum profiles as a function of the opening angle $\theta$ (in radians) in the brick medium (a-1)-(a-2), outward along the expanding medium (b-1)-(b-2), and inward against the expanding medium (c-1)-(c-2). The ratio is taken with respect
to the vacuum results. 
The results for strongly coupled description with (solid) and without particlization (dashed), weakly coupled description (dotted), and in-vacuum (dot-dashed) cases are compared with data of the jet shape ratio between PbPb (0-10\%) and $pp$ collisions at $5.02$ TeV from the CMS collaboration:  
dots for inclusive jets with $p^{\mathrm{jet}}_{T}>120$~GeV \cite{Sirunyan:2018jqr}, and 
square and triangle, respectively, 
for leading and subleading jets in back-to-back dijet events with 
$p^{\mathrm{leading\,jet}}_{T}>120$~GeV and $p^{\mathrm{subleading\,jet}}_{T}>50$~GeV \cite{CMS:2021nhn}. 
}
\label{fig:2}
\end{figure*}

To see detailed structure around the jet direction, 
we also show the jet energy and momentum shapes, 
\begin{eqnarray}
\label{eq:shapes}
\!\!\!\!\!\!\!\!\!
\rho_{E}(\theta)
=\frac{1}{E(\theta\!=\!1)}\frac{dE}{d\theta},\hspace{4pt}
\rho_{P_{x}}(\theta)=\frac{1}{P_{x}(\theta\!=\! 1)}\frac{dP_{x}}{d\theta}, 
\end{eqnarray}
scaled by their vacuum results for the range $0\leq\theta\leq 1.5$ in Fig.~\ref{fig:2}. 
The results for the expanding medium are compared with the experimental data of PbPb/$pp$ jet shape ratios for inclusive jets with $p^{\mathrm{jet}}_{T}>120~\mathrm{GeV}$. 
In addition, comparisons are also made with experimental data for jets that are expected to experience similar background flows 
in back-to-back dijet events: leading jets vs. jets in following flow (outward) and subleading jets vs. jets in opposing flow (intward). 
The modification pattern in the inclusive jet data 
is captured well by the results from our semi-realistic medium, especially in the outward case, which gives the more typical environment in the medium for inclusive jets. 

One can see that the medium flow effects, which cause the unique structures at large angles $\theta \gtrsim \pi/2$, affect the jet inner structure around the jet direction $\theta < 1.5$. 
In the inward case, the area where the backward suppression effect appears is swept more backward by the radial flow, as shown in Fig.~\ref{fig:1}. 
Therefore, in the hemisphere in the jet direction $\theta < 1.5$, the suppression effect itself does not appear, and an overall increase is seen. 
In addition, one can see a flattened structure within $\theta < 1.0$ due to the stretching of the angular distribution of jet energy and momentum by the opposing flow. 
It is worth mentioning that the experimental data for subleading jets shows the same qualitative behavior most prominently. 
In contrast, for the outward case, such a flat region cannot be seen as well as the leading-jet data. 
Instead, the effect of backward suppression 
due to the push by the radial flow starts to appear at $\theta\approx 1$. 
Thus, experimental measurements at larger angles are needed to determine the flow configuration that jets experience more precisely. 

The clear difference between the various scenarios for the dynamics of the medium response to the jet also appears in the large-angle region for the cases with expanding medium. 
The strongly coupled description gives weaker energy and momentum broadening than the weakly coupled case for both following and opposing flows. 
In the outward case with the strongly coupled description, 
even suppression can be seen at $\theta\approx1.5$ due to the interplay between the radial flow and backward over subtraction effects.

In summary, we have systematically studied jet shape modification in a static or an expanding fluid, using a set of very different medium response effects: weakly coupled response, and a strongly coupled response, both with and without particlization. These represent the range of approaches currently applied to the study of medium response to jets. 
The parameters of each approach, as well as those for the expansion of the bulk medium, were chosen to be very close to realistic simulations.

Remarkably, all three approaches lead to very similar modifications of the jet shape structure for typical jet angles of $\theta \lesssim 1$, and these were found to be generally consistent with the observed modification to jet shape distributions.
The differing medium response effects seem to cause differences in the jet shape distribution at very large angles.
This work demonstrates that experimental measurement of jet structure at angles larger than the typical jet angle $\theta \lesssim 1$ is essential for the detailed study of jet medium response. 
On the theory side, modeling QGP medium dynamics is crucial to having realistic flow configurations for precise studies of jet modification in heavy-ion collisions.
Such a precision framework can further shed light on studies applying jets as sources of acoustic probes for the flowing medium. 

\begin{acknowledgments}
The authors are grateful to S. Cao for valuable discussions. 
This work was supported in part by the National Science Foundation (NSF) within the framework of the JETSCAPE collaboration under Grant No.~{ACI-1550300}, by JSPS KAKENHI Grant No.~{22K14041}, and by the U.S. Department of Energy (DOE) under Grant No.~{DE-SC0013460}.
\end{acknowledgments}

\bibliography{ref,ref_hand,for_aps_ref,sm}

\pagebreak

\onecolumngrid

\setcounter{equation}{0}
\setcounter{figure}{0}
\setcounter{table}{0}
\setcounter{page}{1}
\makeatletter

\begin{center}
\textbf{\large Supplemental Material for\\ Bulk medium evolution has considerable effects on jet observables}
\end{center}
\renewcommand{\thetable}{S\arabic{table}}   
\renewcommand{\thefigure}{S\arabic{figure}}
\renewcommand{\theequation}{S\arabic{figure}}

\title{Supplemental Material for\\ Bulk medium evolution has considerable effects on jet observables!}

\author{Yasuki~Tachibana} 
\affiliation{Department of Physics and Astronomy, Wayne State University, Detroit, MI 48201, USA}
\affiliation{
Akita International University, Yuwa, Akita-city 010-1292, Japan
}

\author{Chun~Shen}
\affiliation{Department of Physics and Astronomy, Wayne State University, Detroit, MI 48201, USA}
\affiliation{RIKEN BNL Research Center, Brookhaven National Laboratory, Upton, NY 11973, USA}

\author{Abhijit~Majumder} 
\affiliation{Department of Physics and Astronomy, Wayne State University, Detroit, MI 48201, USA}

\date{\today}

\maketitle

\begin{figure}[htbp]
\begin{center}
\vspace{10pt}
\includegraphics[width=0.7\textwidth,bb=0 0 720 468]{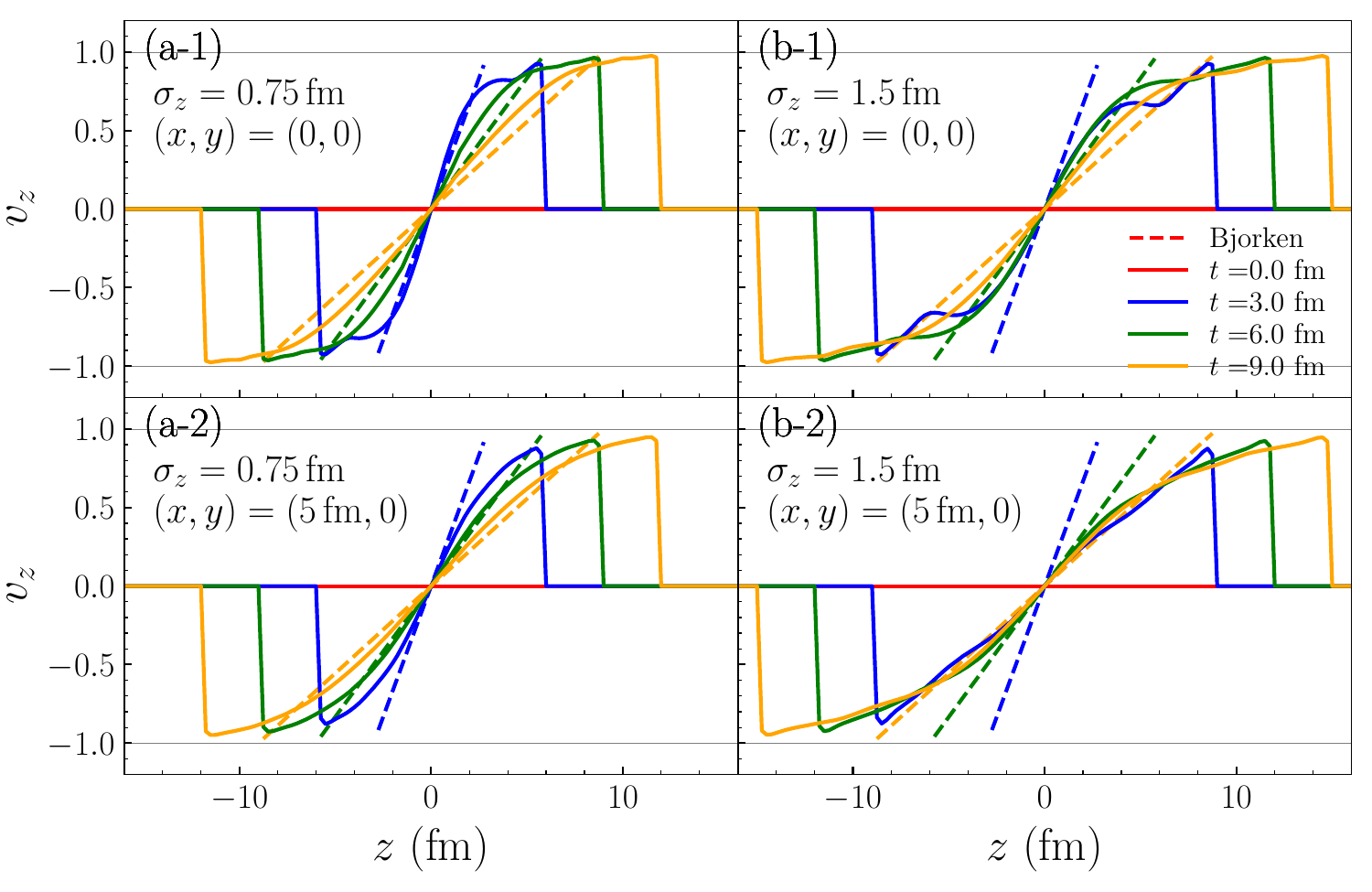}
\vspace{-10pt}
\end{center}
\caption{(Color online) 
Longitudinal flow velocities as a function of $z$ 
at different times $t=0,~3,~6,~9$~fm in the lab frame. 
The transverse coordinates are set to 
$(x,y) = (0,0)$ and $(5\,\mathrm{fm},0)$. 
Solid lines are for the media initialized with 
the oblate Gaussian profile of widths 
$\sigma_{\mathrm{T}}=1.5\,\mathrm{fm}$ and $\sigma_z=0.75\,\mathrm{fm}$~[(a-1) and (a-2)] 
and with the isotropic Gaussian of widths 
$\sigma_{\mathrm{T}}=\sigma_z=1.5\,\mathrm{fm}$~[(b-1) and (b-2)].
Dashed lines represent the Bjorken flow $v_z=z/t$.  
}
\label{fig:zflow}
\end{figure}

In our Letter, 
we employed the expanding medium fluid 
initialized by the oblate three-dimensional Gaussian profile
in Eq. (\ref{eq:profile}). 
The widths of the initial proifle $\sigma_{z}$ and $\sigma_{\mathrm{T}}$ 
are chosen to reproduce a semi-realistic flow configuration in heavy-ion collisions. 
Figure~\ref{fig:zflow} shows 
the medium flow velocity in the $z$ direction $v_z$ 
for the cases with the profile used in the Letter 
(Left plots: $\sigma_{\mathrm{T}}=1.5$~fm and $\sigma_{z}=0.75$~fm) 
and the spherical profile 
(Right plots: $\sigma_{\mathrm{T}}=\sigma_{z}=1.5$~fm), 
compared to Bjorken flow. 
It can be seen that the longitudinal flow in 
our oblate medium is very close to the Bjorken flow at every time step. 
On the contrary, in the spherical medium case, the distribution and evolution of $v_z$ are apparently different from the Bjorken flow: The medium expands slower, especially at an early time. 

\begin{figure*}[htbp]
\begin{center}
\includegraphics[width=0.7\textwidth,bb=0 0 842 709]{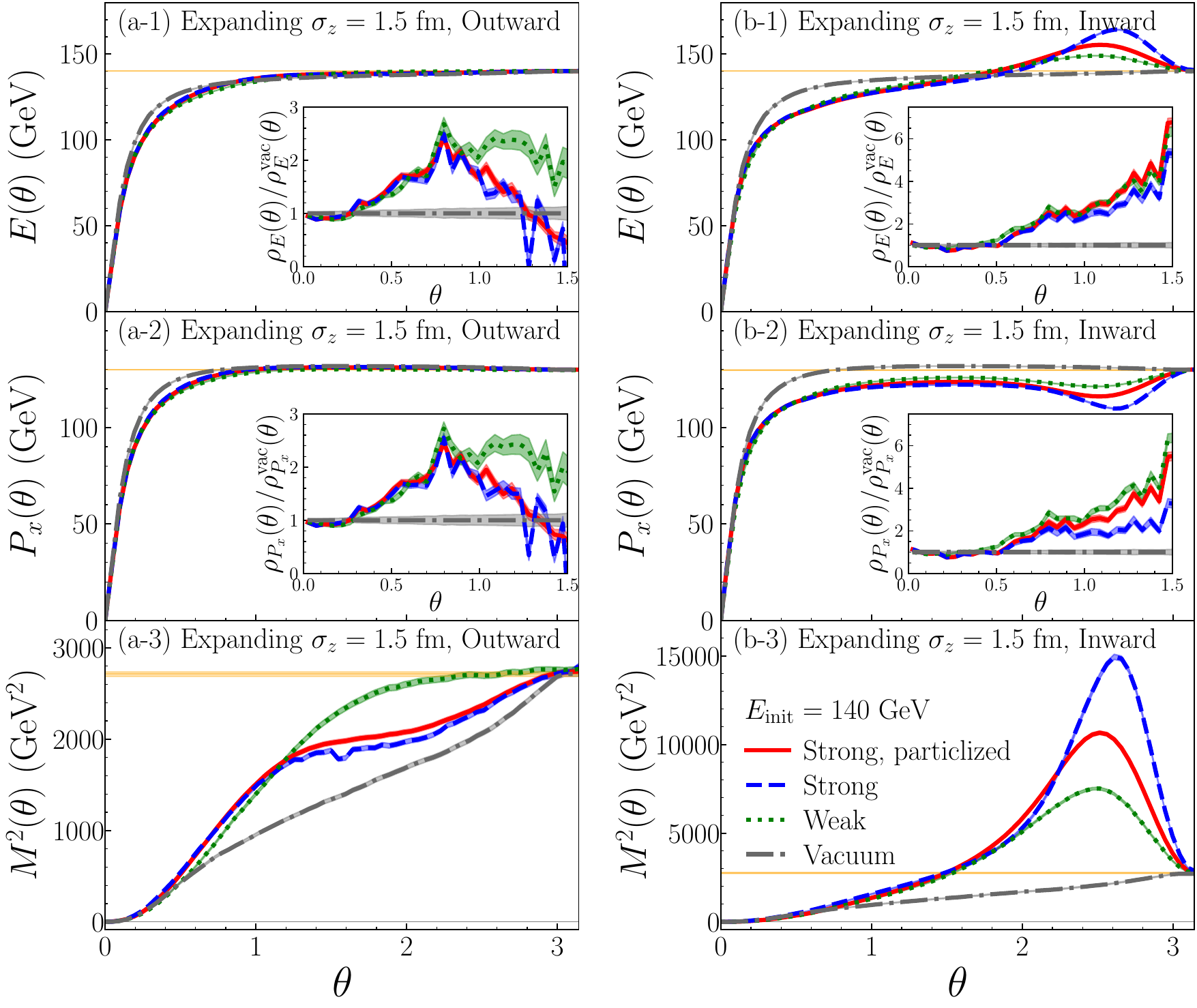}
\vspace{-12pt}
\end{center}
\caption{(Color online) 
Same as Fig.~\ref{fig:1} for jets propagating the expanding medium of $\sigma_z=1.5$~fm with $E_{\mathrm{init}}=140$~GeV. 
The insets show the same ratios of normalized differential quantities as in Fig.~\ref{fig:2}. 
}
\label{fig:140gev_sig150}
\end{figure*}

%
Figure~\ref{fig:140gev_sig150} presents the angular distributions of energy, momentum, and invariant mass 
for jets with $E_{\mathrm{init}}=140$~GeV 
in the spherical expanding medium with $\sigma_{\mathrm{T}}=\sigma_{z}=1.5$~fm. 
The modification patterns are qualitatively the same as in the case of the oblate medium as shown in Figs.~\ref{fig:1} and \ref{fig:2} in the Letter. Meanwhile, the sizes of the modifications are larger with spherical expanding medium for both weakly and strongly coupled scenarios. 
Since the spherical medium expands slower than the oblate one, 
the jet partons interact with the high temperature medium for a longer time. 

\begin{figure*}[htbp]
\vspace{18pt}
\begin{center}
\includegraphics[width=0.98\textwidth,bb=0 0 1278 709]{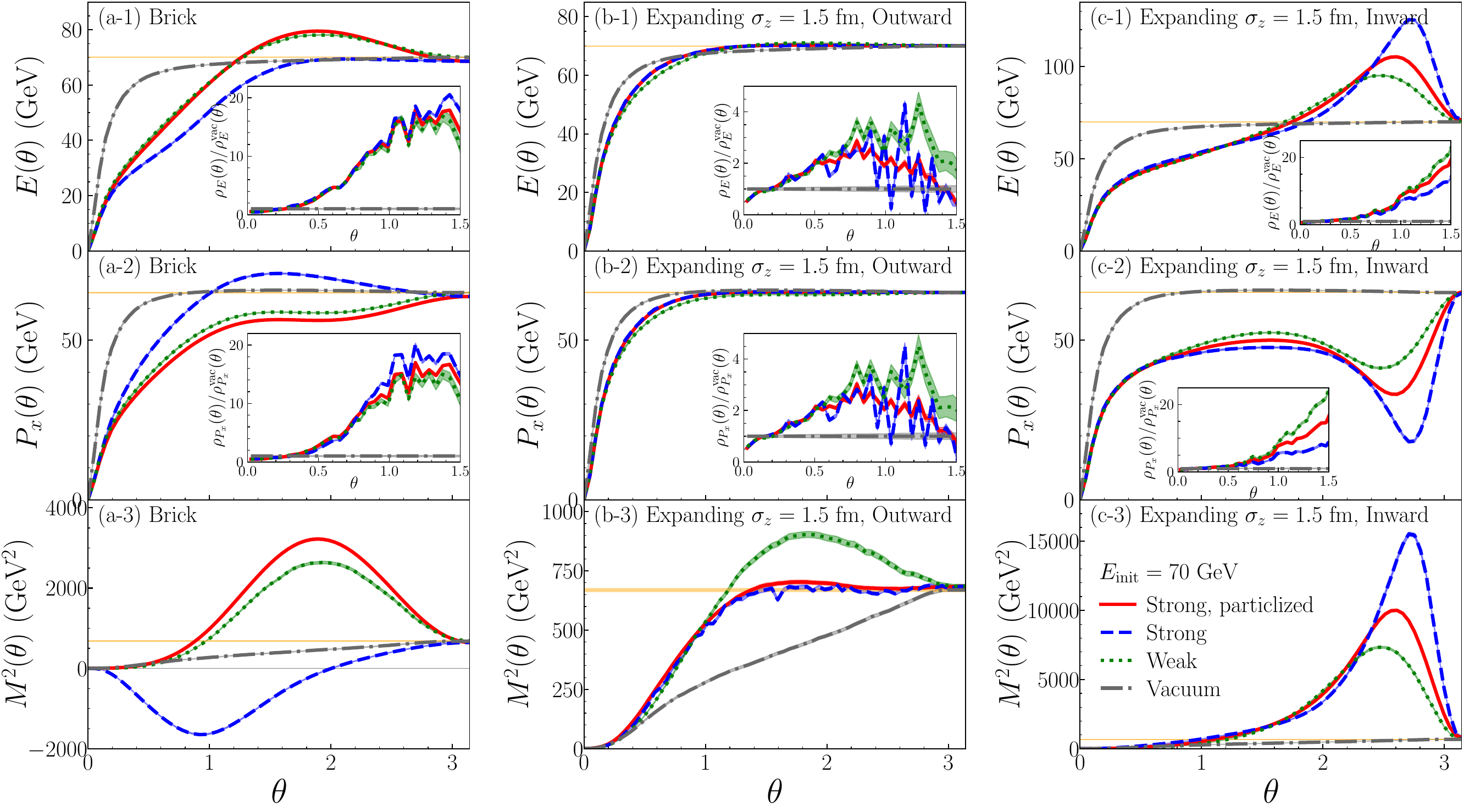}
\vspace{-12pt}
\end{center}
\caption{(Color online) 
Same as Fig. \ref{fig:1} for jets propagating the expanding medium of $\sigma_z=1.5$~fm with $E_{\mathrm{init}}=70$~GeV. 
The insets show the same ratios of normalized differential quantities as in Fig.~\ref{fig:2}. 
}
\label{fig:70gev_sig150}
\end{figure*}

The calculations presented in the Letter are done 
for jets initiated by a parton with energy $E_{\mathrm{init}}=140$ GeV. 
This value is chosen to be close to that of jets triggered with a typical $p^{\mathrm{jet}}_{\mathrm{T}}$ threshold ($\sim 120$ GeV) applied in jet substructure measurements (e.g. Ref.~\cite{Sirunyan:2018jqr}). 
In reality, the jet structure modifications do not qualitatively depend on the initial jet energy. 
Figure \ref{fig:70gev_sig150} shows the same quantities 
as in Fig. \ref{fig:140gev_sig150} but for jets with $E_{\mathrm{init}}=70$ GeV. 
The medium fluid is initialized with the spherical Gaussian profile for the expanding medium cases. 
For all the medium response descriptions 
in the cases of the brick and inward propagation in the expanding medium, 
the modification patterns show almost the same trends as those for $E_{\mathrm{init}}=140$ GeV. 
In the case of the outward propagation in the expanding medium, 
the backward over-subtraction effect, which cannot be seen in the case for $E_{\mathrm{init}}=140$ GeV, 
appears and brings the bump structures in the angular dependence of jet mass squared [Fig.~\ref{fig:70gev_sig150}~(b-3)]. 
This is because partons undergo many scatterings even in the early stages with very small background flow velocity. 
The partons in jets with smaller initial energy have smaller virtualities and 
their virtualities quickly reach the value for the switching to the scattering-dominated phase described by LBT, $\mu=2$~GeV. 
In particular, 
the backward over-subtraction effect is prominent in the weakly-coupled scenario,  
since holes' propagation is not affected by the medium flow at the later stage once they are generated.


\end{document}